\documentclass[sigconf]{acmart}

\AtBeginDocument{%
  }

\copyrightyear{2024}
\acmYear{2024}
\setcopyright{acmlicensed}\acmConference[ESEM '24]{Proceedings of the 18th ACM / IEEE International Symposium on Empirical Software Engineering and Measurement}{October 24--25, 2024}{Barcelona, Spain}
\acmBooktitle{Proceedings of the 18th ACM / IEEE International Symposium on Empirical Software Engineering and Measurement (ESEM '24), October 24--25, 2024, Barcelona, Spain}
\acmDOI{10.1145/3674805.3690746}
\acmISBN{979-8-4007-1047-6/24/10}

\usepackage{listings}
\usepackage{xcolor}

\definecolor{customgray}{HTML}{404040}

\definecolor{codegreen}{rgb}{0,0.6,0}
\definecolor{codegray}{rgb}{0.5,0.5,0.5}
\definecolor{codepurple}{rgb}{0.58,0,0.82}
\definecolor{backcolour}{rgb}{0.95,0.95,0.92}
\lstdefinestyle{code-style}{
    backgroundcolor=\color{backcolour},   
    commentstyle=\color{codegreen},
    keywordstyle=\color{magenta},
    numberstyle=\tiny\color{codegray},
    stringstyle=\color{codepurple},
    basicstyle=\ttfamily\footnotesize,
    breakatwhitespace=false,         
    breaklines=true,                 
    captionpos=b,                    
    keepspaces=true,                 
    numbers=left,                    
    numbersep=5pt,                  
    showspaces=false,                
    showstringspaces=false,
    showtabs=false,                  
    tabsize=2
}

\definecolor{lightergray}{RGB}{240,240,240}

\newcommand{\mybox}[1]{
  \vspace{0.5em}
  \noindent
  \fcolorbox{black}{lightergray}{\parbox{.975\columnwidth}{#1}}
  \vspace{0.5em}
}

\begin{document}

\title{Automatic Library Migration Using Large Language Models: \\ First Results}

\author{Aylton Almeida}
\orcid{0009-0002-0649-856X}
\email{ayltonalmeida@dcc.ufmg.br}
\affiliation{%
  \institution{Federal University of Minas Gerais}
  \city{Belo Horizonte}
  \state{Minas Gerais}
  \country{Brazil}
}

\author{Laerte Xavier}
\orcid{0000-0001-7925-4115}
\email{laertexavier@pucminas.br}
\affiliation{%
  \institution{Pontificial University of Minas Gerais}
  \city{Belo Horizonte}
  \state{Minas Gerais}
  \country{Brazil}
}

\author{Marco Tulio Valente}
\orcid{0000-0002-8180-7548}
\email{mtov@dcc.ufmg.br}
\affiliation{%
  \institution{Federal University of Minas Gerais}
  \city{Belo Horizonte}
  \state{Minas Gerais}
  \country{Brazil}
}

\begin{abstract}
Despite being introduced only a few years ago, Large Language Models (LLMs) are already widely used by developers for code generation. However, their application in automating other Software Engineering activities remains largely unexplored. Thus, in this paper, we report the first results of a study in which we are exploring the use of ChatGPT to support API migration tasks, an important problem that demands manual effort and attention from developers. Specifically, in the paper, we share our initial results involving the use of ChatGPT to migrate a client application to use a newer version of SQLAlchemy, an ORM (Object Relational Mapping) library widely used in Python. We evaluate the use of three types of prompts (Zero-Shot, One-Shot, and Chain Of Thoughts) and show that the best results are achieved by the One-Shot prompt, followed by the Chain Of Thoughts. Particularly, with the One-Shot prompt we were able to successfully migrate all columns of our target application and upgrade its code to use new functionalities enabled by SQLAlchemy's latest version, such as Python's \texttt{asyncio} and \texttt{typing} modules, while preserving the original code behavior.
\end{abstract}



\keywords{API Migration; Large Language Models; ChatGPT; Python; SQL\-Alchemy}


\maketitle

\section{Introduction}

Large Language Models (LLMs) are being used to support several software engineering tasks, including generating tests~\cite{shafer-tests,siddiq-tests,alshahwan-tests}, fixing bugs~\cite{sobania-bugs} and supporting code review and pair programming sessions~\cite{tufano-2021-code-review,tufano-2022-code-review,imai-pair-programming}. However, to the best of our knowledge, they have not yet been used to support API migration. This is usually a key activity in modern software development, as the applications increasingly depend on APIs~\cite{abdalkareem2017reasons}. On the one hand, these APIs typically evolve rapidly to offer new features and increase developers' productivity~\cite{hou2011exploring,lamothe2021systematic,salama2019stability,jss-2018-deprecation}. On the other hand, this evolution often results in the introduction of breaking changes, which are changes in the APIs that impact their clients~\cite{saner2017-laerte,emse2020_aline,bogart2021and,saner2018}. Thus, frequently, current software applications must be updated to use a new and improved API version~\cite{sawant2018reaction}. Normally, this task is manual since we lack established and consolidated tools to support it, at least in production software~\cite{hora2015developers,kula2018developers,wang2020empirical}.

Thus, in this paper, we report the first results of a study in which we are using language models, specifically GPT 4.0, to support the migration of client applications to use newer API versions. Specifically, we chose the SQLAlchemy library as our object of study, which is a very popular Object Relational Mapper (ORM) in the Python ecosystem. This library faced a significant update in its version 2.0 to incorporate Python's typing module. We then chose a real client application and carefully attempted to use language models (particularly, GPT version 4.0) to migrate it to the new version of SQLAlchemy. For this, we explored and evaluate the results of three types of prompts: Zero-Shot (in which we simply describe the task we want the model to perform), One-Shot (where in addition to describing the desired task, we also provide an example of its execution), and Chain Of Thoughts (where we briefly describe the steps necessary for the model to perform the task).

To evaluate the correctness and the quality of the code migrated by GPT we use a set of metrics including number of passing tests and SQLAlchemy-specific metrics, such as number of migrated columns and number of migrated methods. We also consider source code quality metrics, as provided by two popular static analysis tools for Python (Pylint and Pyright). Moreover, we manually migrated the client application to use the newer SQLAlchemy version. Then, we use this version as a ground-truth, particularly for analysing the quality scores generated by the mentioned static analysis tools.
As our last investigation, we evaluate and present the results of the migration of the application's tests.

Our main contributions are twofold: (1) We describe the first version of a framework for library migration using language models, specifically OpenAI's GPT model. The proposed framework consists of a set of prompts and a set of metrics for evaluating the correctness and quality of this task. Although subjected to improvements, we argue this framework can serve as a starting point for developers interested in automating this task. (2) We also present and discuss the first results of using this framework to support the migration of a client of a popular library in the Python ecosystem (SQLAlchemy, version 1 to version 2). We discuss and analyze this migration using the proposed metrics. In addition to the application code, we also perform and analyze the migration of the application's tests.

The remainder of this paper is organized as follows. In Section 2, we describe the study design, detailing the methodology used for the migration process. Section 3 presents the results of the migration, including a comparative analysis of the different prompting methods. We organize this section in two parts. First, in Section 3.1, we present the migration of the application code. Then, in Section 3.2, we assess the migration of the application tests. Section 4 discusses threats to validity and in Section 5 we review related work. Finally, Section 6 concludes the paper and outlines directions for future work.

\section{Study Design}
\label{sec:study-design}
In order to explore the effectiveness of LLMs to support API migration, we carefully attempt to use GPT 4.0 to upgrade the SQLAlchemy API in a client application. 
In this section, we detail the steps taken to conduct this migration, describing the context of this task and the creation and evaluation of three types of prompts.

\subsection{Target API and Client Application}
\label{sec:target-library-and-client-application}


SQLAlchemy\footnote{https://github.com/sqlalchemy/sqlalchemy} is an Object-Relational Mapping (ORM) API to support the integration of Python applications with relational databases, abstracting operations related to database connection and manipulation. It supports the connection to various databases, such as PostgreSQL, MySQL, and Oracle, using a simple API that resembles pure SQL queries. The API is well-known and -adopted in the open source community, with more than 8,9K stars, 1,3K forks, and 765K users among other GitHub repositories.

In this paper, we focus on migrating from version 1 to version 2 of SQLAlchemy. Version 1 is widely used and provides all the functionalities expected from a robust ORM. However, as new features have been added to Python, SQLAlchemy was updated to take advantage of such improvements. Among the major changes in version 2, the compatibility with Python’s static typing stands out, improving error detection during development and facilitating the maintenance of large applications. While compatibility with Python’s \texttt{asyncio} was introduced in version 1.4, version 2.0 includes several performance improvements and optimizations for asynchronous operations. Particularly, the \texttt{asyncio} module supports efficient asynchronous tasks using the \texttt{async/await} syntax in Python. Finally, SQLAlchemy 2 has improved the query syntax, making it more intuitive and easier to use.

To perform the migration of the API, we used a client application called \texttt{BiteStreams/fastapi-template}. It is a Python application that uses the FastAPI library, a popular web framework for creating APIs, in conjunction with the SQLAlchemy ORM. The application implements a TODO list feature, containing REST routes for creating and listing tasks, which are stored in a PostgreSQL database. It contains a single table called \texttt{todo} with four columns, responsible for storing information about each todo item in the list. Additionally, it has 18 methods, which allows the user to fetch a single TODO item, list all of them, and insert new ones into the database. We selected this client application because it includes four automated tests, including both integration and unit tests, facilitating the verification of code behavior and functionality after migration. Although it is a simple project, we claim that it provides an interesting environment to perform our analysis.



\subsection{Migration Process}

\begin{figure}[!h]
\begin{lstlisting}[
        language=Python,
        caption={Example of manually migrated code},
        label={list:migrated-code}
    ]
@lru_cache(maxsize=None)
def get_engine(db_string: str):
    return create_async_engine(db_string, pool_pre_ping=True, poolclass=NullPool)


class TodoInDB(SQL_BASE):  # type: ignore
    __tablename__ = "todo"

    id: Mapped[int] = mapped_column(Integer, primary_key=True, autoincrement=True)
    key: Mapped[str] = mapped_column(String(length=128), nullable=False, unique=True)
    value: Mapped[str] = mapped_column(String(length=128), nullable=False)
    done: Mapped[bool] = mapped_column(Boolean, default=False)
\end{lstlisting}
\end{figure}

In order to establish a baseline for the results, we started by performing a manual migration of the API in the client application. Listing \ref{list:migrated-code} presents an excerpt of the migrated code. This manual migration allowed us to identify particular changes required and the potential challenges that GPT could find during the migration process. Additionally, it helped us to understand some adjustments needed to ensure that the automated migration process would work. For example, during this initial exploration, we observed the necessity of upgrading the library version in the dependencies file and adding both \texttt{asyncpg} and \texttt{pytest-asyncio} APIs to the project. Such APIs support the connection with the database and the execution of tests when using Python's \texttt{asyncio}.

After the manual migration of the client application, we decided to break the automatic migration process in two steps. The first step consists on migrating only the application code with the assistance of GPT. For this reason, we excluded the test files from the prompts. In other words, the tests were manually migrated to assess the code produced by GPT. 
In the second step, we tasked GPT to migrate only the test files. In this case, our intention is to assess the migrated tests by running them on the manually migrated application. In both steps, we evaluated three types of prompts, as we will describe in Section~\ref{sec:prompts}.

For the GPT migration, we implemented a simple script to transition between prompting approaches. 
It uses OpenAI’s Python API version 1.14 together with the Chat Completions API. We used model GPT-4 and a basic role was set up for the system: "You are a developer with expertise in Python", along with a temperature of zero, which should result in more consistent outputs. For each prompt, the first result was stored for analysis, since we noticed that running the same prompt multiple times yields slightly different results, even though the temperature setting was set to zero.

\subsection{Definition of Prompts}
\label{sec:prompts}

\begin{figure}[!h]
  \centering
  \fcolorbox{gray!60}{lightergray}{%
    \parbox{\dimexpr\linewidth-2\fboxsep-2\fboxrule\relax}{%
      \noindent\makebox[\linewidth]{\colorbox{customgray}{\parbox{\dimexpr\linewidth-2\fboxsep-2\fboxrule\relax}{\textbf{\color{white}Base Prompt}}}}\\[0.2cm]
      The Python code bellow uses the library sqlalchemy with version 1. Migrate it so that it works with version 2 of sqlalchemy. Make the code compatible with python's \texttt{asyncio}. Use python's typing module to add type hints to the code. Your answer must only contain code. Do not explain it. Do not add markdown backticks for code. Do not add extra functionality to the code. Do not remove code that is not being changed. If there's no need to change the code, answer only with the code itself. The first line of code must have a comment "\#\#\# START CODE \#\#\#". The last line of code must have a comment "\#\#\# END CODE \#\#\#". \\[-0.15cm]

      Here is the code to migrate: \\[-0.15cm]
    
      \#\#\# START CODE \#\#\# \\
      ...
    }%
  }
  \caption{Base command used for all prompts}
  \label{fig:base-prompt}
\end{figure}

A critical part of working with LLMs is defining prompts that clearly communicate the intentions of the user. In this study, we explore three different prompting methods. The first is a Zero-Shot approach, where the model receives a task description in the prompt but no example is provided to illustrate the expected output~\cite{sahoo2024systematic}. The second prompt uses a One-Shot approach, providing the model with an output example to help it understand the given task. Lastly, a Chain Of Thoughts approach was used. In this case, the prompt contains a step-by-step guide that lead to the final output~\cite{wei2023chainofthought}. The same base command was used for the three prompts, as shown in Figure \ref{fig:base-prompt}. In this command, we instruct the model to return only code, with no explanations. This is relevant to better automate the task. Additionally, we ask it to use both Python’s \texttt{asyncio} and typing features in order to ensure it adopts the newest features and syntax for SQLAlchemy 2.

The first prompting method used was the Zero-Shot approach. In this method, no code examples are passed, giving the LLM complete freedom on how to perform its task. It is composed only by the base command and the code to be migrated, as already presented in Figure~\ref{fig:base-prompt}. The second method was a One-Shot approach, in which a code example is included with the prompt to give the model a reference for the required migration. This can be seen in Figure \ref{fig:One-Shot-prompt}. The provided example was retrieved from the examples folder in the SQLAlchemy repository, which defines some tables and operations using the newest library version, together with \texttt{asyncio} and typing. 

\begin{figure}[!h]
  \centering
  \fcolorbox{gray!60}{lightergray}{%
    \parbox{\dimexpr\linewidth-2\fboxsep-2\fboxrule\relax}{%
      \noindent\makebox[\linewidth]{\colorbox{customgray}{\parbox{\dimexpr\linewidth-2\fboxsep-2\fboxrule\relax}{\textbf{\color{white}One-Shot Prompt}}}}\\[0.2cm]
      Use the following code block as an example of migrated code, follow the same patterns used in it: \\[-0.15cm]
      
      \#\#\# START CODE \#\#\# \\
      ...
    }%
  }
  \caption{One-Shot prompt}
  \label{fig:One-Shot-prompt}
\end{figure}

Lastly, a Chain Of Thoughts prompt was used, where a step-by-step guide is included along with the example to instruct the migration process. The added step-by-step guide can be seen in Figure \ref{fig:cot-prompt}. It was meant to provide GPT a guide on how to proceed with the migration process, detailing actions such as using an async engine and applying the new column declaration methods.

\begin{figure}[!h]
  \centering
  \fcolorbox{gray!60}{lightergray}{%
    \parbox{\dimexpr\linewidth-2\fboxsep-2\fboxrule\relax}{%
      \noindent\makebox[\linewidth]{\colorbox{customgray}{\parbox{\dimexpr\linewidth-2\fboxsep-2\fboxrule\relax}{\textbf{\color{white}Chain of Thoughts Prompt}}}}\\[0.2cm]
      Use the steps bellow as a guide for the migration. You don't need to follow them exactly as described, but they should be able to help with the migration: \\[-0.15cm]

    1. Update the used database engine, if any, so that you're using `create\_async\_engine` instead of `create\_engine`. \\
    2. If any tables and their columns are declared, update their declarations so that they use `mapped\_columns` instead of `schema.Column` and ensure they are correctly typed with the Mapped annotation, making sure to import the correct types from the library. \\
    3. Ensure that all queries, if any, are updated to use the new 2.0 style of querying, such as using `select()` instead of `query()`. \\
    4. Update functions that use `sessionmaker` to use `session` instead. \\
    5. Update the code to use async functions and await calls where necessary. \\
    6. Implement type hinting for all functions and variables and update old type hinting to ensure they are correct. \\
    7. Ensure there are no missing import statements. \\
    8. Remove any unused imports or variable declarations. \\
    9. Make sure the code works. \\
    ...
    }%
  }
  \caption{Chain of Thoughts prompt}
  \label{fig:cot-prompt}
\end{figure}

\begin{table*}[!t]
    \centering
    \caption{Application Migration Results}
    \begin{tabular}{lcccccc}
        \toprule
        Metrics & \textbf{Runs Successfully} & \textbf{Tests} & \textbf{Pylint} & \textbf{Pyright} & \textbf{Migrated Columns} & \textbf{Migrated Methods} \\
        \midrule
        Before Migration    & Yes   & 4/4   & 7.68/10   & 2 errors  & -     & -     \\
        Manual Migration    & Yes   & 4/4   & 7.97/10   & 0 errors  & 4/4   & 18/18 \\ \midrule
        Zero-Shot           & No    & 0/4   & 7.30/10   & 6 errors  & 0/4   & 16/18 \\
        One-Shot            & Yes   & 4/4   & 7.77/10   & 7 errors  & 4/4   & 13/18 \\
        Chain Of Thoughts   & No    & 0/4   & 7.33/10   & 5 errors  & 4/4   & 17/18 \\
        \bottomrule
    \end{tabular}
    \label{tab:only-application-prompt-results}
\end{table*}

\subsection{Metrics}
To evaluate the effectiveness of the migration process and assess whether the migrated application works as intended, we used the following metrics:

\begin{itemize}
    \item \textit{Runs successfully:} Checks whether the code runs without crashing after the migration.
    
    \item \textit{Number of tests that pass:} The number of tests that pass after the migration process. This aims to check if the application still works as expected after the migration. In total, 4 out of 4 tests should pass correctly.
    
    \item \textit{Pylint score:} This tool is a commonly used linter in Python. A base configuration file was created, and after the code was migrated, the linter was run in order to check for common errors, such as missing imports or unused variables.
    
    \item \textit{Pyright score:} This tool is a type checker commonly used in Python. It was run to search for possible typing errors in the code. It gives the number of errors and warnings found.
    
    \item \textit{Number of migrated columns:} The number of columns that were correctly migrated to the new syntax. In total, there should be 4 migrated columns.
    
    \item \textit{Number of migrated methods:} This is the number of functions and methods that were migrated to use typing and \texttt{asyncio} in their signature. In total, 18 of them needed to be updated in order to correctly use both features.
    
    \item \textit{Number of migrated tests:} This is the number of tests correctly migrated. A test is considered migrated when it uses the new features necessary for it to run correctly after the rest of the code has been migrated, such as Python's \texttt{async} annotation. There are 4 tests that need migration.
\end{itemize}

\section{Results}
In this section, we present and analyze the results achieved by ChatGPT using the three prompts defined in the study. First, we present the results for the application migration (Section~\ref{sec:results-application}) and then we also discuss the migration of the tests of this application (Section~\ref{sec:results-tests}).

\subsection{Application Migration}
\label{sec:results-application}

Table~\ref{tab:only-application-prompt-results} presents the results for migrating the application using Zero-Shot, One-Shot, and Chain Of Thoughts prompts. To facilitate comparison and analysis, we also present the results for the original application (before migration) and for the baseline migration, i.e., the one we performed manually. In this first migration step, we only migrate the application code, as mentioned in Section \ref{sec:target-library-and-client-application}. Moreover, we use the manually migrated tests to evaluate the functionality of the code produced by the LLM.\footnote{Just to clarity, the migration of the tests will be discussed in Section 3.2.}\\

\noindent \textbf{Zero-Shot Prompt:} When analyzing the results for the Zero-Shot approach, we can see in Table \ref{tab:only-application-prompt-results} that it did not perform well. In fact, we could not run the application after the migration. This was mostly due to errors that prevented the application code from opening a connection with the database. This problem can be seen in more detail in Listing \ref{list:Zero-Shot-import-errors}. First, in line 2 an attempt was made to import the \texttt{create\_async\_engine} method from the wrong module. This method is implemented in the module \texttt{sqlalchemy.ext. asyncio}, but the automatically migrated code attempts to import it from \texttt{sqlalchemy}. Additionally, in line 4, ChatGPT tried importing a method called \texttt{create\_async\_session}, but this method does not exist. Due to these errors, Pylint score was lower when compared to the original application (7.30 vs 7.86, respectively).

\begin{figure}[!h]
\begin{lstlisting}[
        language=Python,
        caption={Code generated by the zero-shot prompt with errors in \texttt{imports}},
        label={list:Zero-Shot-import-errors}
    ]
from pydantic import BaseModel
from sqlalchemy import Boolean, Column, Integer, String, create_async_engine
from sqlalchemy.exc import DatabaseError
from sqlalchemy.ext.asyncio import AsyncSession, create_async_session
from sqlalchemy.orm import declarative_base, sessionmaker
from sqlalchemy.pool import NullPool
\end{lstlisting}
\end{figure}

Another factor that resulted in a poor result for the Zero-Shot prompting was its inability to use Python’s typing feature. This can be seen in lines 4-7 in Listing \ref{list:Zero-Shot-class-definition}, where the columns were not updated to use the new \texttt{mapped\_column} method. Moreover, the columns (\texttt{id}, \texttt{key}, \texttt{value}, and \texttt{done}) were not declared using type information.

\begin{figure}[!h]
\begin{lstlisting}[
        language=Python,
        caption={Code generated by the Zero-Shot prompt missing types},
        label={list:Zero-Shot-class-definition}
    ]
class TodoInDB(SQL_BASE):  # type: ignore
    __tablename__ = "todo"

    id = Column(Integer, primary_key=True, autoincrement=True)
    key = Column(String(length=128), nullable=False, unique=True)
    value = Column(String(length=128), nullable=False)
    done = Column(Boolean, default=False)
\end{lstlisting}
\end{figure}

Finally, in Listing \ref{list:Zero-Shot-typing-errors}, we can see other typing errors, such as using an \texttt{Iterator} instead of an \texttt{AsyncIterator} as the return type for the \texttt{create\_todo\_repository} function (lines 2-3). Due to such problems, the number of errors raised by Pyright was higher when compared with the one of the original implementation (6 vs 2 type errors, respectively).

\begin{figure}[!h]
\begin{lstlisting}[
        language=Python,
        caption={Code generated by the Zero-Shot using incorrect types},
        label={list:Zero-Shot-typing-errors}
    ]
@asynccontextmanager
async def create_todo_repository() -> Iterator[TodoRepository]:
    async with create_async_session(get_engine(os.getenv("DB_STRING"))) as session:
        todo_repository = SQLTodoRepository(session)

        try:
            yield todo_repository
        except Exception:
            await session.rollback()
            raise
        finally:
            await session.close()
\end{lstlisting}
\end{figure}

\noindent \textbf{One-Shot Prompt:} In contrast with the Zero-Shot approach, the One-Shot method yielded significantly better results. The migrated code runs as expected, and all tests also passed. It is interesting to see that when we provided an example to the GPT model, all columns and methods were migrated correctly, and there were no import errors. A fragment of the migrated code can be seen in Listing \ref{list:One-Shot-class-definition}. As we can observe, ChatGPT was able to correctly add types to the table columns and use the \texttt{mapped\_column} method to map them. For a better understanding, the reader can compare the code in Listing \ref{list:One-Shot-class-definition} with the one presented in Listing \ref{list:Zero-Shot-class-definition}, which was generated using Zero-Shot prompt.

\begin{figure}[!h]
\begin{lstlisting}[
        language=Python,
        caption={Code generated by the One-Shot prompt with correct types},
        label={list:One-Shot-class-definition}
    ]
class TodoInDB(SQL_BASE):  # type: ignore
    __tablename__ = "todo"

    id = Column(Integer, primary_key=True, autoincrement=True)
    key = Column(String(length=128), nullable=False, unique=True)
    value = Column(String(length=128), nullable=False)
    done = Column(Boolean, default=False)
\end{lstlisting}
\end{figure}

As a negative note, both Pylint and Pyright scores decreased when the application was migrated using an One-Shot prompt. Two unused imports were generated, which resulted in a lower Pylint score. Additionally, a relevant typing error was detected in the migrated code, where instead of typing the \texttt{session} attribute as an \texttt{AsyncSession}, it was typed as a \texttt{Session}, resulting in multiple errors throughout the application.\\

\noindent \textbf{Chain Of Thoughts Prompt:} This prompt also performed well, resulting in the correct migration of both methods and columns, similar to the One-Shot prompt. It also achieved the second-best Pylint results (after the One-Shot approach), and obtained the best Pyright result, with five errors compared to 6 and 7 errors for the Zero-Shot and One-Shot prompts, respectively. However, the \texttt{create\_async\_engine} was imported from the wrong module, resulting in an error that prevented the application from running and the tests from passing. We claim this is a minor error that can be fixed manually by a developer with experience in Python, thus allowing the migrated application to execute and behave as expected.\\

\mybox{{\em Summary:} (1) One-Shot was the prompt with the best results: it was able to generate a running application that passes the tests; it also achieve the best Pylint score.  (2) Chain Of Thoughts was the second best prompt and the one with the lowest number of Pyright type errors (five errors); the code did not execute due to a minor import error. (3) Zero-Shot presented the worst results and it was not able to correctly migrate any of the table columns.}

\subsection{Tests Migration}
\label{sec:results-tests}

Table \ref{tab:tests-prompt-results} presents the results for the three prompts when migrating the tests and running them against the manually migrated application. The column Migrated Tests shows the number of tests correctly migrated. An example of a correctly migrated test can be seen in Listing \ref{list:migrated-test}. In this example, a \texttt{test\_todo} was implemented using an \texttt{async} function from Python's \texttt{asyncio} module (line 2). Also, throughout the test both \texttt{await} and \texttt{async} keywords were inserted when necessary to ensure a correct behavior in asynchronous calls  (lines 3, 4 and 6). 

\begin{table}[!h]
    \centering
    \caption{Tests Migration Results}
    \begin{tabular}{lcc}
        \toprule
        Metrics & \textbf{Migrated Tests} & \textbf{Tests that Pass} \\
        \midrule
        Before Migration    & --  & --  \\
        Manual Migration    & 4/4 & 4/4 \\ \midrule
        Zero-Shot           & 4/4 & 1/4 \\
        One-Shot            & 4/4 & 1/4 \\
        Chain Of Thoughts   & 4/4 & 1/4 \\
        \bottomrule
    \end{tabular}
    \label{tab:tests-prompt-results}
\end{table}

\begin{figure}[!h]
\begin{lstlisting}[
        language=Python,
        caption={Correctly Migrated Test},
        label={list:migrated-test}
    ]
@pytest.mark.integration
async def test_todo(todo_repository: SQLTodoRepository):
    async with todo_repository as r:
        await r.save(Todo(key="testkey", value="testvalue"))

    todo = await r.get_by_key("testkey")
    assert todo.value == "testvalue"
\end{lstlisting}
\end{figure}

However, when looking at Table \ref{tab:tests-prompt-results}, we can see that although migrated correctly only a single test has initially passed. This was due to an issue in the way the tests are implemented, which added a new layer of complexity to the migration process. Essentially, the tests are implemented in such a way that between their execution a fixture must run (in Python, a fixture is a function that runs before each test). In our case, this function truncates the TODOs table and initializes a new session to ensure the database is empty before running the next test. However, none of the approaches were able to correctly migrate this fixture. Therefore, after the first test ran and passed, the other three tests failed due to duplicate key errors when trying to insert the same row as the previous test in the table. 

It is interesting to point out that all three approaches managed to migrate the tests in the same way, including making the same mistakes while migrating the mentioned fixture. This incorrect implementation can be seen in Listing \ref{list:migrated-fixture}. In this code, we can see that an attempt to truncate the table occurs in lines 13 and 14. However, there is no call to the \texttt{commit} function, which causes the session to close without applying the \texttt{truncate} command. Before SQLAlchemy version 2, all sessions had an \texttt{autocommit=True} behavior, such that when they were closed, they automatically committed the changes. Nevertheless, in version 2, this behavior changed to \texttt{autocommit=False}, which resulted in the problem for all three prompts and explains why ChatGPT made this kind of mistake. Although this error prevented the tests from passing, it can be easily fixed by an experienced developer. Indeed, we applied this fix and then all the four tests started to pass.\\

\begin{figure}[!t]
\begin{lstlisting}[
        language=Python,
        caption={Migrated Fixture},
        label={list:migrated-fixture}
    ]
@pytest.fixture
async def todo_repository() -> SQLTodoRepository:
    time.sleep(1)
    alembicArgs = ["--raiseerr", "upgrade", "head"]
    alembic.config.main(argv=alembicArgs)

    engine = create_async_engine(os.getenv("DB_STRING"))
    async_session = sessionmaker(engine, class_=AsyncSession)

    async with async_session() as session:
        yield SQLTodoRepository(session)

        await session.execute(
            ";".join([f"TRUNCATE TABLE {t} CASCADE" for t in SQL_BASE.metadata.tables.keys()])
        )
\end{lstlisting}
\end{figure}

\mybox{{\em Summary:} An interesting issue occurred during the migration of the tests. Initially, they were migrated correctly. However, a subtle change introduced in the new version of SQLAlchemy prevented the migrated tests from executing successfully. Specifically, in the new library version, \texttt{autocommit} is no longer true by default. Thus, once we manually restored the expected commit behavior, the tests migrated with the three prompts passing. }

\section{Threats to Validity}

The first threat is related to the selection of only one application as target client for migration. We acknowledge that different applications may demand different approaches to migrate. This means that other applications may yield different results when applying the same methodology. Besides, we also highlight that we used our framework only with Python and the SQLAlchemy API. Changing both the programming language and the target API may  produce different observations.

We also acknowledge that the GPT model usually provides different answers in each interaction, even with the same prompt. 
To mitigate this threat, we have set the \texttt{temperature} parameter to zero, which makes the model more deterministic. We also relied on the first answer of each interaction, as reported in Section~\ref{sec:study-design}. Lastly, the way the prompts have been constructed may also influence the outputs. 
In this case, we carefully defined our three types of prompts, getting inspired by the ones used in the literature.

\section{Related Work}

Using LLMs to support software development has been previously explored in the literature. Several studies have been conducted to understand how to take advantage of their potential to save development time and improve code quality. For example, Schäfer et al.~\cite{shafer-tests} explored the usage of LLMs for the implementation of unit tests. In their study, ChatGPT was used to implement unit tests for 25 different npm packages. Their experiment suggests that prompting the LLM with a Few-Shot approach positively influences the results. This conclusion is aligned with the findings of our work. By providing an example for the migration (as in the One-Shot and Chain Of Thoughts approaches), GPT was able to support an improved migration compared to the Zero-Shot prompting.

Sobania et al.~\cite{sobania-bugs} assess ChatGPT's bug-fixing capabilities using a benchmark called QuixBugs. The authors find that, even though ChatGPT was not built specifically for code repair, it is extremely competitive with other approaches created for this purpose, such as CoCoNut and Codex. Due to its interface, users can provide extra information about the problems, such as error messages and expected outputs, which further increase the chances of success. In this work, we focus on code migration instead of bug fixing, but we also achieve results that suggest that GPT can also be used to support code maintenance tasks.

\section{Conclusion and Future Work}

In this paper, we proposed a LLM-based framework for upgrading API versions using three different prompting approaches and a set of metrics to asses the migrated code. We use this framework to migrate a client application to work with a newer version of the SQLAlchemy library. We concluded that LLMs are able to correctly migrate the project when provided with at least one example of an already migrated application, making only minor mistakes such as inserting unused imports or incorrectly typed variables, which are errors than can be later fixed by a developer.

Although our first results are promising, there is still future work to be done in order to use LLMs to support library migrations. Thus, we plan to continue to work as follows:

\begin{enumerate}
    \item We intend to evaluate other libraries, including libraries for other programming languages, such as Java (an established statically-typed programming language) and JavaScript (a popular language in the specific domain of Web apps).
    
    \item We plan to define and evaluate other types of prompts (e.g., Few-Shots and Chain Of Symbols) and improve the current prompts. For example, in the case of Chain of Thoughts we can add more details on the step-by-step guide or even pass the official migration documentation for the LLM to use as a reference.
    
    \item We plan to assess our framework with developers, for example, by submitting pull requests in GitHub projects.
    
    \item Finally, we also intend to evaluate other LLMs, such as Google Gemini\footnote{https://gemini.google.com} and Amazon Q.\footnote{https://aws.amazon.com/pt/q}
\end{enumerate}

\noindent{\bf Replication Data:} The code of the application used in this research, along with the code of all migrated versions, is available at: \url{https://zenodo.org/records/11403035}

\begin{acks}
This research was supported by grants from CNPq and FAPEMIG.
\end{acks}

\bibliographystyle{ACM-Reference-Format}
\bibliography{references}

\end{document}